\documentclass[journal=jctcce,manuscript=article,layout=traditional]{achemso}

\pdftrailerid{}

\usepackage[utf8]{inputenc}
\usepackage[T1]{fontenc}
\usepackage{mathtools}
\usepackage{graphicx}
\usepackage{lineno}
\usepackage{xcolor}
\usepackage{hyperref}

\usepackage{mathptmx}
\usepackage{etoolbox}
\usepackage[section]{placeins}
\usepackage{soul}
\usepackage[version=4]{mhchem}
\usepackage{threeparttable}
\usepackage{booktabs}
\usepackage{multirow}
\usepackage{longtable}
\usepackage{makecell}

\usepackage{amsmath}
\usepackage{amsfonts}
\usepackage{amssymb,amsthm,esint}
\usepackage{fullpage}
\usepackage{bm}
\usepackage{hyperref}
\usepackage{textcomp}
\usepackage{bigints}
\usepackage{dsfont}
\usepackage{colortbl}
\usepackage[normalem]{ulem}
\usepackage{empheq}
\usepackage{dsfont}
\usepackage{comment}
\usepackage{enumitem}
\usepackage[colorinlistoftodos]{todonotes}

\makeatletter
\newcommand{\warninput}[1]{\filename@parse{#1}\InputIfFileExists{#1}{}{\message{LaTeX Warning: File `\filename@base.\ifx\filename@ext\relax tex\else\filename@ext\fi'not found on input line \the\inputlineno}}}
\makeatother

\DeclareMathAlphabet\mathbfcal{OMS}{cmsy}{b}{n}

\DeclareMathOperator*{\argmin}{arg\,min}

\newcommand{\rhoa}{\rho_a}

 \def\R{\mathbb{R}}

\newcommand{\br}{\mathbf r}

\newcommand{\bR}{\mathbf R}

\usepackage{relsize}

\newcommand{\m}[1]{\texttt{#1}}

\newcommand{\clsna}[1]{\textbf{a#1±}}

\newcommand{\clspa}[1]{\textbf{a#1+}}

\title{Approximations of the Iterative Stockholder Analysis scheme using exponential basis functions}

\author{YingXing Cheng}
\affiliation[University of Stuttgart]{Institute of Applied Analysis and Numerical Simulation, University of Stuttgart, Pfaffenwaldring 57, 70569, Stuttgart, Germany}

\author{Benjamin Stamm}
\affiliation[University of Stuttgart]{Institute of Applied Analysis and Numerical Simulation, University of Stuttgart, Pfaffenwaldring 57, 70569, Stuttgart, Germany}
\email{benjamin.stamm@mathematik.uni-stuttgart.de}

\begin{document}

    \begin{abstract}
        In this work, we introduce several approximations of the Iterative Stockholder Analysis (ISA) method based on exponential basis functions.
These approximations are categorized into linear and non-linear models, referred to as LISA and NLIS, respectively.
By particular choices of hyperparameters in the NLIS model, both LISA and the Minimal-Basis Iterative Stockholder (MBIS) method can be reproduced.
Four LISA variants are constructed using systematically generated exponential basis functions derived from the NLIS model applied to atomic systems.
The performance of these LISA variants and NLIS models is benchmarked on 15 small molecules, including neutral, anionic, and cationic species.
To facilitate comparison, we propose several metrics designed to highlight differences between the methods.
Our results demonstrate that LISA, employing Gaussian basis functions derived from the NLIS model on isolated atomic systems, achieves an optimal balance of computational accuracy, robustness, and efficiency, particularly in minimizing the objective function.
     \end{abstract}

    \maketitle

    \newpage

    \section{Introduction}
    \label{sec:introduction}
    Defining an atom in a multi-atom molecule remains an open question in chemistry.
This is because the partitioning of molecular orbitals or electron density into atomic contributions is not an intrinsic property in quantum mechanics.
Consequently, different partitioning methods can yield significantly different atom-in-molecule (AIM) properties, such as atomic charges and multipole moments, which are crucial for developing molecular polarizable force fields.\cite{Warshel1976,Kaminski2002,Ren2002,Jorgensen2007,Shi2013,Lamoureux2003a,Yu2003,Lemkul2016,Kunz2009,Mortier1985,Mortier1986,Kale2011,Kale2012,Kale2012a,Bai2017,Cools-Ceuppens2022,Verstraelen2013,Verstraelen2014,Cheng2022}
The classification and comparison of these methods have been discussed extensively in the recent studies~\cite{Ayers2000,C.Lillestolen2008,Lillestolen2009,Bultinck2009,Verstraelen2012b,Verstraelen2013a,Misquitta2014b,Heidar-Zadeh2015,Verstraelen2016,Heidar-Zadeh2018,Benda2022} and references therein.

In this work, we focus on the Iterative Stockholder Analysis (ISA) family of methods.\cite{C.Lillestolen2008,Lillestolen2009}
The ISA method employs the Kullback–Leibler (KL) divergence as an objective functional to minimize the dissimilarity between the AIM electron density and the reference proatomic electron densities, subject to the constraint that the sum of the proatomic densities, also known as the promolecular density, equals the molecular electron density.
In the original ISA method,\cite{C.Lillestolen2008,Lillestolen2009} the proatomic density corresponds to the spherical average of the AIM density.

Although the solution of the ISA method is analytically unique and convex,\cite{Benda2022} it is not numerically robust, and the spherical average of the AIM or proatomic density does not monotonically decay in certain systems, which is counterintuitive in the chemistry community.\cite{Verstraelen2012b}
To address these numerical challenges, Verstraelen \textit{et al.} developed an ISA variant known as GISA.
In this approach, the proatomic density is approximated using a set of normalized $s$-type Gaussian basis functions with unknown non-negative expansion coefficients.\cite{Verstraelen2012b}
These coefficients are optimized via least-squares fitting to the proatomic density from ISA, subject to the constraint that the population of the AIM density equals that of the proatomic density.
However, the quadratic objective function in GISA leads to over-delocalized proatomic densities, resulting in unphysical charges.
Furthermore, atomic charges derived from GISA are not highly transferable, and conformational stability is low.\cite{Heidar-Zadeh2018}

To resolve the issue of unphysical proatomic density tails, Misquitta \textit{et al.} proposed a variant, BS-ISA, that combines Gaussian functions for short-range contributions with Slater functions for long-range behavior.
Subsequently, Verstraelen \textit{et al.} introduced another variant, MBIS, which uses a minimal set of Slater functions with unknown non-negative expansion coefficients and corresponding exponents.\cite{Verstraelen2016}
The MBIS method also employs the KL divergence as the objective function, subject to constraints similar to GISA.
While MBIS significantly improves upon GISA and performs well in chemical applications, excessively negative atomic charges are observed in molecular anions, likely due to the limited number of Slater basis functions.\cite{Heidar-Zadeh2018}

Another ISA variant, the Additive Variational Hirshfeld (AVH) model, recently proposed by Farnaz \textit{et al.}, employs densities from selected states of isolated atoms and ions as basis functions.\cite{Heidar-Zadeh2017,Heidar-Zadeh2024}
Unlike other ISA variants, AVH basis functions are not analytic.
Later, Benda \textit{et al.} proposed a unified framework for all ISA variants.
These methods can be classified as linear or nonlinear ISA approximations (LISA and NLIS, respectively) based on whether nonlinear parameters are included in the basis functions to be optimized.
LISA models, such as ISA, GISA, and AVH, exhibit advantageous mathematical properties, including uniqueness, convexity, and guaranteed convergence, while NLIS models, such as BS-ISA and MBIS, lack these guarantees.

A new LISA model with Gaussian basis functions similar to GISA has also been proposed in Ref.~\citenum{Benda2022}, with the key difference being the use of KL divergence for optimizing expansion coefficients.
Recently, we investigated the numerical performance of LISA models and found that LISA with Gaussian basis functions achieves lower divergence compared to GISA and MBIS, except for certain molecular anions with hydrogen atoms where MBIS performs better.
This can be addressed by incorporating more diffuse Gaussian functions for hydrogen atoms in LISA, although its suitability for chemical applications remains an open question.
A systematic approach to constructing basis functions is still needed, and this work aims to address this gap.

In this work, we focus exclusively on ISA variants employing exponential basis functions.
We derive an NLIS model by treating expansion coefficients, exponents, and the order of the exponential basis functions as unknown parameters.
The order $p$ of an exponential basis function, defined as $g_{ak}(r) = c_{ak}\,e^{-\alpha_{ak}r^p}$, distinguishes between a Slater-type decay ($p = 1$) and a Gaussian-type decay ($p = 2$).
The resulting stationary equations resemble those in MBIS but include an additional equation for the order of the exponential basis functions.
By imposing further simplifications, we propose several NLIS models, including a reproduction of the MBIS model by fixing the order to one and the number of basis functions to the number of rows for each element in the periodic table.
Additionally, we propose a generalized MBIS model using exponential basis functions with orders ranging from one to two and the same number of basis functions as in MBIS.

Furthermore, we introduce LISA models combining Gaussian and/or Slater functions, with the exponents of exponential basis functions determined using NLIS models to fit atomic data.
These models are benchmarked using 15 small molecules, including anions, cations, and neutral molecules.
Four boron wheel molecules with delocalized bonding environments are also included.

In Section~\ref{sec:methods}, we describe the relevant methodology for this study.
Section~\ref{sec:details} provides computational details.
Results and discussions are presented in Section~\ref{sec:res}, and a summary is given in Section~\ref{sec:summary}.
Atomic units are used throughout.

    \section{Methods}
    \label{sec:methods}
    \subsection{ISA partitioning methods}
The information loss when AIM densities are approximated by proatomic functions is modeled using the KL divergence,
\begin{align}
    s_\text{KL}(\rhoa | \rhoa^0) &= \int_{\R^3} \rhoa(\br) \ln \left( \frac{\rhoa(\br)}{\rhoa^0(\br)} \right) d\br,
    \label{eq:atomic_divergence}
\end{align}
using the conventions
\begin{align}
    0\cdot \ln\left(\frac{0}{0}\right)&=0,                   &
    p\cdot\ln\left(\frac{p}{0}\right) &=\infty \quad \forall p>0, &
    0\cdot\ln\left(\frac{0}{p}\right) &=0,
    \label{eq:conventions}
\end{align}
where $\rhoa^0(\br)$ and $\rhoa(\br)$ are the proatom and AIM densities of atom $a$, respectively.
Thus, the information loss when AIM densities are approximated by proatoms in a molecule with $N_\text{atoms}$ atoms can be expressed as the sum of KL divergences for each atom:
\begin{align}
    S(\{\rhoa\}, \{\rhoa^0\}) = \sum_{a=1}^{N_\text{atoms}} s_\text{KL}(\rhoa|\rhoa^0).
    \label{eq:divergence}
\end{align}

In practice, several constraints are imposed to ensure that the partitioning results align with chemical and physical properties.
First, the sum of AIM densities must equal the total density $\rho(\br)$:
\begin{align}
    \sum_{a=1}^{N_\text{atoms}} \rhoa(\br) = \rho(\br).
\end{align}
Second, the population of each proatom and its corresponding AIM density should be equal, avoiding ambiguity in the statistical interpretation of Eq.~\eqref{eq:divergence}:\cite{Parr2005,Verstraelen2016}
\begin{align}
    N_a = \int_{\R^3} \rhoa(\br) d\br = \int_{\R^3} \rhoa^0(\br) d\br.
    \label{eq:pop_cond}
\end{align}

The Lagrangian for the problem is defined as follows:
\begin{align}
    L[\{\rhoa\}, \lambda(\br), \{\rhoa^0\}]
    =
    & \sum_{a=1}^{N_{\text{atoms}}} \int_{\R^3} \rhoa(\br) \ln \left( \frac{\rhoa(\br)}{\rhoa^0(\br)} \right) \, d\br \nonumber \\
    & + \int_{\R^3} \lambda(\br) \left( \sum_{a=1}^{N_{\text{atoms}}} \rhoa(\br) - \rho(\br) \right) d\br \nonumber \\
    &+ \sum_{a=1}^{N_{\text{atoms}}} \mu_a \left( \int_{\R^3}  (\rhoa^0(\br) - \rhoa(\br)) d\br \right),
    \label{eq:lagrangian_original}
\end{align}
where $\lambda(\br)$ is the Lagrange multiplier associated with the first constraint, and $\mu_a$ are Lagrange multipliers associated with the consistency between the proatom and AIM populations.
Solving the Euler-Lagrange equations of Eq.~\eqref{eq:lagrangian_original} leads to the well-known stockholder partitioning formula:
\begin{align}
    \rhoa(\br) &= \rho(\br) \frac{\rhoa^0(\br)}{\rho^0(\br)}, \label{eq:stockholder} \\
    \rho^0(\br) &= \sum_b \rho_b^0(\br),
\end{align}
where $\rho^0(\br)$ is also referred to as the pro-molecule density.

\subsection{Exponential-type basis functions}
Different choices of basis functions to discretize $\rho^0_a(\br)$ lead to different variants of the ISA method.
In this work, we introduce $m_a$ positive basis functions $g_{a,k}$ for each site $\bR_a$, centered at $\bR_a$ and radially symmetric.
Although $\bR_a$ can represent an arbitrary expansion center, as pointed out in Ref.~\citenum{Benda2022}, this work focuses on cases where $\bR_a$ denotes the position of the nucleus with index $a$.
Moreover, we assume, without loss of generality, that $g_{a,k}(\br)$ is normalized such that
\begin{align}
    \int_{\R^3} g_{a,k}(\br) \, d\br=1,
    \label{eq:normalization}
\end{align}
and we focus on exponential functions of the following form:
\begin{align}
  g_{a,k}(\br) = \frac{n_{a,k} \alpha_{a,k}^{3/n_{a,k}}}{4 \pi \Gamma(3/n_{a,k})}  e^{-\alpha_{a,k} |\br - \bR_a|^{n_{a,k}}}.
  \label{eq:general_exp_basis_func}
\end{align}
Here, $a$ and $k$ denote the indices of atoms and basis functions, respectively.
For Gaussian basis functions, there holds $n_{a,k} = 2$ for all $a$ and $k$, while for Slater basis functions, $n_{a,k} = 1$, corresponding to
\begin{align}
    g_{a,k}(\br) = \left(\frac{\alpha_{a,k}}{\pi}\right)^3 e^{-\alpha_{a,k} |\br-\bR_a|^2},
\end{align}
and
\begin{align}
    g_{a,k}(\br) = \frac{\alpha_{a,k}^3}{8\pi} e^{-\alpha_{a,k} |\br-\bR_a|},
\end{align}
respectively.

\subsection{General Lagrangian formulation}
By discretizing the proatom densities using general exponential basis functions, the discretized version of the Lagrangian in Eq.~\eqref{eq:lagrangian_original} can be rewritten as presented in Ref.~\citenum{Verstraelen2016}:
\begin{align}
    L[\{\rhoa\}, \lambda(\br), \{c_{a,k}\}, \{\alpha_{a,k}\}, \{n_{a,k}\}, \{\mu_a\}]
    =
    & \sum_{a=1}^{N_{\text{atoms}}} \int_{\R^3} \rhoa(\br) \ln \left( \frac{\rhoa(\br)}{\rhoa^0(\br)} \right) \, d\br \nonumber \\
    & + \int_{\R^3} \lambda(\br) \left( \sum_{a=1}^{N_{\text{atoms}}} \rhoa(\br) - \rho(\br) \right) d\br \nonumber \\
    &+ \sum_{a=1}^{N_{\text{atoms}}} \mu_a \left( \int_{\R^3}  (\rhoa^0(\br) - \rhoa(\br)) d\br \right),
    \label{eq:lagrangian}
\end{align}
with
\begin{align}
    \rhoa^0(\br) &= \sum_k \rho_{a,k}^0(\br), \\
    \rho_{a,k}^0(\br) &= c_{a,k} g_{a,k}(\br).
\end{align}

Independent variation of the Lagrangian $L$ with respect to each variable $(\rhoa(\br), c_{a,k}, \alpha_{a_k}, n_{a_k})$ leads to a set of Euler-Lagrange equations, which, together with the constraints, determine the AIM and proatom densities, as well as the Lagrange multipliers $\lambda(\br)$ and $\{\mu_a\}$.

\subsubsection{Derivative with respect to $c_{a,k}$}
First, consider the derivative of $L$ with respect to $c_{a,k}$:\cite{Verstraelen2016}
\begin{align}
    0 = \frac{\partial L}{\partial c_{a,k}}
    & = \int_{\R^3}  \frac{\delta L}{\delta \rhoa^0(\br)} \frac{\partial \rhoa^0(\br)}{\partial c_{a,k}} \, d\br \\
    & = \int_{\R^3}  \left( -\frac{\rhoa(\br)}{\rhoa^0(\br)} + \mu_a \right) g_{a,k}(\br) d\br \\
    & = \mu_a - \int_{\R^3}  \frac{\rhoa(\br)}{\rhoa^0(\br)} g_{a,k}(\br) d\br.
\end{align}
Multiplying by $c_{a,k}$ and summing over the shells $k$ of atom $a$, using Eq.~\eqref{eq:pop_cond}, we obtain,
\begin{align}
    0 = \sum_k c_{a,k} \frac{\partial L}{\partial c_{a,k}}
    & = \mu_a N_a^0 - \int_{\R^3}  \frac{\rhoa(\br)}{\rhoa^0(\br)} \sum_k c_{a,k} g_{a,k}(\br) d\br \\
    & = \mu_a N_a^0 - N_a,
\end{align}
where
\begin{align}
    N_a^0 = \sum_k c_{a,k},
\end{align}
with $N_a = N_a^0$ due to Eq.~\eqref{eq:pop_cond}, and $\mu_a = 1$ for each atom.

\subsubsection{Functional derivative with respect to $\rhoa(\br)$}
Next, the functional derivative of $L$ with respect to $\rhoa(\br)$ is given by,\cite{Verstraelen2016}
\begin{align}
    0 = \frac{\delta L}{\delta \rhoa(\br)} &= \ln \left( \frac{\rhoa(\br)}{\rhoa^0(\br)} \right) + \lambda(\br),
\end{align}
which leads to the stockholder partitioning formula,
\begin{align}
    \rhoa(\br) &= \rho(\br) \frac{\rhoa^0(\br)}{\rho^0(\br)}.
    \label{eq:aim_density}
\end{align}

\subsubsection{Derivative with respect to $\alpha_{a,k}$}
Now, consider the derivative of $L$ with respect to $\alpha_{a,k}$,
\begin{align}
    0 = \frac{\partial L}{\partial \alpha_{a,k}}
    & = \int_{\R^3}  \frac{\delta L}{\delta \rhoa^0(\br)} \frac{\partial \rhoa^0(\br)}{\partial \alpha_{a,k}} d\br \\
    & = -\int_{\R^3}  \frac{\rhoa(\br)}{\rhoa^0(\br)} \frac{\partial \rhoa^0(\br)}{\partial \alpha_{a,k}}
      + \mu_a c_{a,k} \frac{\partial}{\partial \alpha_{a,k}} \int_{\R^3}  g_{a,k}(\br) d\br \\
    & = \int_{\R^3}  \frac{\rhoa(\br)}{\rhoa^0(\br)} \left( \frac{3}{n_{a,k} \alpha_{a,k}} - |\br - \bR_{a}|^{n_{a,k}} \right) \rho_{a,k}^0(\br) d\br,
\end{align}
where Eq.~\eqref{eq:normalization} is used.

\subsubsection{Derivative with respect to $n_{a,k}$}
Finally, consider the derivative of $L$ with respect to $n_{a,k}$,
\begin{align}
    0 = \frac{\partial L}{\partial n_{a,k}}
    & = \int_{\R^3}  \frac{\delta L}{\delta \rhoa^0(\br)} \frac{\partial \rhoa^0(\br)}{\partial n_{a,k}} d\br \\
    & = -\int_{\R^3}  \frac{\rhoa(\br)}{\rhoa^0(\br)} \frac{\partial \rhoa^0(\br)}{\partial n_{a,k}}
      + \mu_a c_{a,k} \frac{\partial}{n_{a,k}} \int_{\R^3}  g_{a,k}(\br) d\br \\
    & = \int_{\R^3}  \frac{\rhoa(\br)}{\rhoa^0(\br)} \left( \frac{1}{n_{a,k}} - \frac{3 \ln(\alpha_{a,k})}{n_{a,k}^2} - \alpha_{a,k} |\br - \bR_a|^{n_{a,k}} \ln({|\br - \bR_a|}) + \frac{3 \psi(3/n_{a,k})}{n_{a,k}^2} \right) \rho_{a,k}^0(\br) d\br
\end{align}
with $\psi(z) = \frac{\Gamma'(z)}{\Gamma(z)}$, where Eq.~\eqref{eq:normalization} is also used.

\subsubsection{Stationary points}
The Lagrangian defined in Eq.~\eqref{eq:lagrangian} can therefore be simplified as\cite{Heidar-Zadeh2015,Verstraelen2016,Heidar-Zadeh2024}
\begin{align}
    L[\{\rhoa\}, \{c_{a,k}\}, \{\alpha_{a,k}\}, \{n_{a,k}\}]
    =
    \sum_{a=1}^{N_{\text{atoms}}} \int_{\R^3} \rhoa(\br) \ln \left( \frac{\rhoa(\br)}{\rhoa^0(\br)} \right) \, d\br
    + \sum_{a=1}^{N_{\text{atoms}}} \left( \int_{\R^3}  (\rhoa^0(\br) - \rhoa(\br)) d\br \right),
    \label{eq:lagrangian_loc}
\end{align}
such that it leads to exactly the same Euler–Lagrange Equations.

Indeed, the corresponding stationary points for $c_{a,k}$, $\alpha_{a,k}$, and $n_{a,k}$ are given by solving the following Euler-Lagrange equations,
\begin{align}
    c_{a,k} &= \int_{\R^3} \frac{\rhoa(\br)}{\rhoa^0(\br)} \rho_{a,k}^0(\br) d\br,
    \label{eq:stationary_points_c_ak}
    \\
\alpha_{a,k} &= \frac{3 c_{a,k}}{n_{a,k}} \left(\int_{\R^3} \frac{\rhoa(\br)}{\rhoa^0(\br)} \rho_{a,k}^0(\br) |\br - \bR_a|^{n_{a_k}} d\br \right)^{-1},
    \label{eq:stationary_point_alpha_ak}
    \\
n_{a,k} &= 3 \ln(\alpha_{a,k}) - 3 \psi(3/n_{a_k})
            + n_{a_k}^2 \alpha_{a,k} \int_{\R^3}  \frac{\rhoa(\br)}{\rhoa^0(\br)} |\br - \bR_a|^{n_{a_k}} \ln({|\br - \bR_a|}) g_{a_k}(\br) d\br,
    \label{eq:stationary_points_n_ak}
\end{align}
in conjunction with Eq.~\eqref{eq:aim_density}.

\subsection{ISA variants}
A general solver for Eqs.~\eqref{eq:stationary_points_c_ak}--\eqref{eq:stationary_points_n_ak} is the self-consistent method proposed in Ref.~\citenum{Verstraelen2016}.
In this section, different ISA variants are reproduced by fixing the variables defined in Eq.~\eqref{eq:lagrangian} using exponential basis functions as given in Eq.~\eqref{eq:general_exp_basis_func}.
More efficient solvers can then be employed depending on the corresponding variant.

The first variant is the nonlinear approximation of the ISA methods, referred to as NLIS,\cite{Benda2022} in which $c_{a,k}$ and $\alpha_{a,k}$ are treated as variables, while $n_{a,k}$ is specified in advance as a hyperparameter.
The reason $n_{a,k}$ is not treated as a variable is due to numerical difficulties, such as instabilities, that arise when solving Eq.~\eqref{eq:stationary_points_n_ak}, which involves the computation of higher-order moments.

When $n_{a,k}=1$ is fixed and $k$ corresponds to the number of subshells of the corresponding neutral atoms, the MBIS method is reproduced.
A more generalized method compared to MBIS uses hyperparameters $n_{a,k}$, which are specified in advance.
This method is referred to as the generalized MBIS method (GMBIS) in this work.

The self-consistent solver (\texttt{SC}) defined in Refs.~\citenum{Verstraelen2016,Cheng2024} is employed in this work to solve the Euler-Lagrange equations for all nonlinear models, including NLIS, MBIS, and GMBIS.
It should be noted that the solution to NLIS is generally not unique.
However, with a good initial guess, a chemically reasonable solution can typically be obtained.\cite{Verstraelen2016}

Furthermore, if both $n_{a,k}$ and $\alpha_{a,k}$ are fixed, i.e., only $c_{a,k}$ is treated as a variable, the linear approximation of the ISA method, denoted LISA, is reproduced.\cite{Benda2022,Cheng2024}
It should be noted that only exponential basis functions are considered in this work; in principle, any valid basis functions can be used.
For example, one could also use the spherical average of atomic densities of atoms with partial charges, and the AVH method would be reproduced.\cite{Heidar-Zadeh2024}

Different solvers can be applied in the LISA model, and based on whether $c_{a,k}$ is allowed to be negative, these solvers can be classified into two categories: \clsna{LISA} (allowing negative $c_{a,k}$) and \clspa{LISA} (using only non-negative $c_{a,k}$).\cite{Cheng2024}
Different solvers within each category yield the same solutions because the optimization problem in the LISA model is convex.
Therefore, we use \texttt{SC} and \texttt{M-NEWTON} as representatives of \clspa{LISA} and \clsna{LISA}, respectively.
Further details can be found in Ref.~\citenum{Cheng2024}.

\subsection{Numerical computation}
Analytically, solving the Euler-Lagrange equations derived from Eq.~\eqref{eq:lagrangian_loc} is equivalent to solving those derived from an alternative extended Lagrangian:\cite{Verstraelen2016,Cheng2024}
\begin{align}
    L_\text{glob}[\{c_{a,k}\}, \{\alpha_{a,k}\}, \{n_{a,k}\}]
    = \int_{\R^3} \rho(\br) \ln\left( \frac{\rho(\br)}{\rho^0(\br)} \right) d\br
    + \int_{\R^3} \left(\rho^0(\br) - \rho(\br)\right) d\br,
    \label{eq:lagrangian_glob}
\end{align}
in conjunction with Eq.~\eqref{eq:lagrangian_loc}, where the AIM densities $\rhoa(\br)$ are given as a set of the proatom densities $\rhoa^0(\br)$.
Note that Eq.~\eqref{eq:lagrangian_glob} is independent of the AIM densities $\rhoa(\br)$ and depends only on the molecular density.
However, methods derived from Eq.~\eqref{eq:lagrangian_loc} and Eq.~\eqref{eq:lagrangian_glob} are generally not equivalent when using numerical quadrature, as the results depend on the grids used.

In principle, one can use both molecular grids and atomic grids for numerical integration.
A molecular grid is constructed from a set of atomic grids, each defined as the tensor product of a radial grid and a spherical grid.
Furthermore, nuclear weights, e.g., the Becke weight function ($w^\text{B}(\br)$), are employed to avoid double-counting in grid points located in the overlap region of two atomic grids.
The numerical version of Eq.~\eqref{eq:lagrangian_glob} on a molecular grid can thus be written as:
\begin{align}
    L_\text{glob}[\{c_{a,k}\}, \{\alpha_{a,k}\}, \{n_{a,k}\}]
    & = \sum_{i=1}^{N_\text{pt}^\text{M}} w^\text{M}(\br_i) \rho(\br_i) \ln\left( \frac{\rho(\br_i)}{\rho^0(\br_i)} \right)
    + \sum_{i=1}^{N_\text{pt}^\text{M}} w^\text{M}(\br_i) \left(\rho^0(\br_i) - \rho(\br_i)\right)
    \nonumber \\
& = \sum_{a=1}^{N_\text{atom}} \sum_{j=1}^{N_\text{pt}^{\text{A}a}} w^{\text{A}a}(\br_{aj}) w^\text{B}(\br_{aj}) \rho(\br_{aj}) \ln\left( \frac{\rho(\br_{aj})}{\rho^0(\br_{aj})} \right)
    \nonumber \\
    & \quad + \sum_{a=1}^{N_\text{atom}} \sum_{j=1}^{N_\text{pt}^{\text{A}a}} w^{\text{A}a}(\br_{aj}) w^\text{B}(\br_{aj}) \left(\rho^0(\br_{aj}) - \rho(\br_{aj})\right),
    \label{eq:lagrangian_glob_num}
\end{align}
with
\begin{align}
    w^\text{M}(\br_i) &= w^{\text{A}a}(\br_{aj}) w^\text{B}(\br_{aj}), \\
    N_\text{pt}^\text{M} &= \sum_{a=1}^{N_\text{atom}} N_\text{pt}^{\text{A}a},
\end{align}
where $w^\text{M}(\br_i)$ is the weight of the molecular grid at point $i$, $w^{\text{A}a}(\br_{aj})$ is the weight of the atomic grid $\text{A}a$ of atom $a$ at point $aj$, and $w^\text{B}(\br_{aj})$ is the Becke weight at point $aj$.
Here, the single index $i$ is used for the molecular grid, while the double indices $aj$ are used for atomic grids, with each $i$ corresponding to a unique $aj$.
$N_\text{pt}^\text{M}$ denotes the number of molecular grid points, and $N_\text{pt}^{\text{A}a}$ represents the number of grid points in the atomic grid $\text{A}a$ of atom $a$.

Using atomic grids, the numerical version of Eq.~\eqref{eq:lagrangian} can be written as:
\begin{align}
    L[\{c_{a,k}\}, \{\alpha_{a,k}\}, \{n_{a,k}\}]
    &= \sum_{a=1}^{N_\text{atom}} \sum_{j=1}^{N_\text{pt}^{\text{A}a}} w^{\text{A}a}(\br_{aj}) w_a(\br_{aj}) \rho(\br_{aj}) \ln\left( \frac{\rho(\br_{aj})}{\rho^0(\br_{aj})} \right) \nonumber \\
    & \quad + \sum_{a=1}^{N_\text{atom}} \sum_{j=1}^{N_\text{pt}^{\text{A}a}} w^{\text{A}a}(\br_{aj}) w_a(\br_{aj}) \left(\rho^0(\br_{aj}) - \rho(\br_{aj})\right),
    \label{eq:lagrangian_num}
\end{align}
where $w_a(\br_{aj})$ is the AIM density weight of atom $a$ at point $aj$.
Since Eq.~\eqref{eq:lagrangian_num} is independent of the nuclear weights, we use Eq.~\eqref{eq:lagrangian_num} with only atomic grids, rather than the molecular grid, for all methods in this work.
Therefore, the difference between Eq.~\eqref{eq:lagrangian_glob_num} and Eq.~\eqref{eq:lagrangian_num} is to use the Becke weights or the Stockholder partitioning as partition of unity to localize the integrals.

\subsection{Functionals and quantities of interest for ISA variants}
In this section, several quantities of interest are introduced to assess the differences among the ISA variants.
First, we decompose the atomic divergence defined in Eq.~\eqref{eq:atomic_divergence} into two components:
\begin{align}
    s_\text{KL}(\rhoa | \rhoa^0) = s_a^\text{I+II} = s_a^\text{I} + s_a^\text{II},
    \label{eq:atomic_divergence_two_parts}
\end{align}
where
\begin{align}
    s_a^\text{I} = s_\text{KL}(\rhoa | \langle \rhoa \rangle_s) &= \int_{\R^3} \rhoa(\br) \ln \left( \frac{\rhoa(\br)}{\langle \rhoa \rangle_s(r) } \right) d\br,
    \label{eq:atomic_divergence_part_1}
\end{align}
and
\begin{align}
    s_a^\text{II} = s_\text{KL}(\langle \rhoa \rangle_s | \rhoa^0) &= \int_{\R^3} \langle \rhoa \rangle_s(r) \ln \left( \frac{\langle \rhoa \rangle_s(r) }{\rhoa^0(\br)} \right) d\br  \nonumber \\
    &= 4\pi \int r^2 \langle \rhoa \rangle_s(r) \ln \left( \frac{\langle \rhoa \rangle_s(r)}{\rhoa^0(r)} \right) dr,
    \label{eq:atomic_divergence_part_2}
\end{align}
with the spherical average of the atomic density $\langle \rhoa \rangle_s(r)$ defined as:
\begin{align}
    \langle \rhoa \rangle_s(r) = \frac{1}{4\pi} \int_0^{\pi} \sin(\theta) \int_0^{2\pi} \rhoa(\br) d\theta d\phi.
    \label{eq:rhoa_sph_ave}
\end{align}

Here, $s_a^\text{I}$ corresponds to the divergence arising from the anisotropy of the AIM density of atom $a$ in the molecule.
If the atomic density is spherically symmetric, $s_a^\text{I}$ is zero; otherwise, it is nonzero.
On the other hand, $s_a^\text{II}$ represents the divergence due to the completeness of the basis functions used to construct the proatom density of atom $a$.
In the complete basis set limit, there holds $s_a^\text{II} = 0$.

We also introduce two quantities:
\begin{align}
    n_a(r) &= 4\pi r^2 \langle \rhoa \rangle_s(r), \label{eq:n_a_r}\\
    s_a(r) &= 4\pi r^2 \langle \rhoa \rangle_s(r) \ln \left( \frac{\langle \rhoa \rangle_s(r)}{\rhoa^0(r)} \right),
    \label{eq:s_a_r}
\end{align}
where $n_a(r) dr$ represents the contribution to the atomic population of atom $a$ in the interval $[r, r+dr]$.\cite{Verstraelen2012b}
Similarly, $s_a(r) dr$ represents the contribution to the atomic divergence of atom $a$ in the interval $[r, r+dr]$, revealing discrepancies between $\langle \rhoa \rangle_s(r)$ and $\rhoa^0(r)$.

Furthermore, we introduce the absolute KL divergence, denoted as $t_\text{KL}$.
The corresponding atomic absolute KL divergence, $t_a$, is then defined to describe the deviation between $\rhoa(\br)$ and $\rhoa^0(\br)$, following the conventions in Eq.~\eqref{eq:conventions}:
\begin{align}
    t_a = t_\text{KL}(\rhoa|\rhoa^0)
    &= \int_{\R^3} \rhoa(\br) \left| \ln \left(\frac{\rhoa(\br)}{\rhoa^0(\br)}\right) \right| d\br,
    \label{eq:t_a_I_and_II}
\end{align}
and the molecular absolute KL divergence of a moleucle is then defined as
\begin{align}
    T = \sum_a t_a.
    \label{eq:mol_abs_kl}
\end{align}

The purpose of introducing the absolute KL divergence $T$ is to provide a more robust measure of the overall discrepancy between $\rhoa^0(\br)$ and their reference counterparts $\rhoa(\br)$.
While the standard KL divergence is sensitive to both the magnitude and sign of deviations, it may yield deceptively small values when large positive and negative contributions cancel each other out.
For example, this can occur when $\langle \rhoa \rangle_s(r) > \rhoa^0(r) > 0$ and $\langle \rhoa \rangle_s(r) < \rhoa^0(r) < 0$ in Eq.~\eqref{eq:s_a_r}, respectively.
In contrast, $T$ captures the cumulative magnitude of deviations without permitting such cancellations, thereby providing a clearer measure of the total error in the density reconstruction.

    \section{Computational details}
    \label{sec:details}
    As introduced in Section~\ref{sec:methods}, the implementation of the ISA method and its variants requires numerical quadrature, with atomic grids employed.
Each atomic grid consists of radial and angular components.
After convergence testing, 150 Gauss-Chebyshev radial points and 770 Lebedev-Laikov angular points were used for all atoms.
The Gauss-Chebyshev integration interval $[-1, 1]$ was mapped to the semi-infinite radial interval $[0, +\infty)$, following Ref.~\citenum{Becke1988}.

All calculations were performed using the \texttt{Horton-Part 1.1.7} package.\cite{YingXing2023}
The input files were prepared using the \texttt{Horton 3} package.\cite{Chan2024,Kim2024,Tehrani2024,Denspart2024}
The \texttt{IOData}~\cite{Verstraelen2021} and \texttt{GBasis}~\cite{Kim2024} packages were used to prepare the molecular density and its gradients on grid points generated by the \texttt{Grid} package.\cite{Tehrani2024}

For benchmarking, five molecules (\ce{CCl4}, \ce{CS2}, \ce{CH4}, \ce{NH3}, \ce{H2O}) and six charged molecular ions (\ce{CH3+}, \ce{CH3-}, \ce{NH4+}, \ce{NH2-}, \ce{H3O+}, \ce{OH-}) were taken from Ref.~\citenum{Cheng2024}.
Molecular structures were optimized using DFT at the B3LYP/aug-cc-pVDZ level of theory.
Additionally, four boron wheel molecules were included: \ce{B8}, \ce{B8^{2-}}, \ce{B8} triplet, and \ce{B9-}.
The structures of \ce{B8}, \ce{B8} triplet, and \ce{B8^{2-}} are perfect heptagons, while \ce{B9-} forms a perfect octagon.
These structures were obtained from Ref.~\citenum{Zhai2003}.
The optimized coordinates of all test molecules are provided in Tables S1--S15 in Section S1 in the Supporting Information.

To compare the numerical performance of different partitioning methods, molecular density calculations were performed at the B3LYP/aug-cc-pVTZ level for all 15 molecules.
The corresponding results are presented in Section~\ref{ssec:numerical}.
Additionally, to assess the sensitivity of different partitioning methods to basis sets and computational approaches, various exchange-correlation (XC) functionals and basis sets were tested for \ce{CH3+}, \ce{CH4}, \ce{CH3-}, \ce{NH4+}, \ce{NH3}, \ce{NH2-}, \ce{H3O+}, \ce{H2O}, and \ce{OH-}.
Specifically, restricted HF, LDA, PBE, and B3LYP functionals were employed in combination with the (d-aug)-cc-pVXZ basis sets, where X = D, T, Q, and 5, resulting in a total of 12 basis sets.
The corresponding results are provided in Section~\ref{ssec:chemical}.

In addition, isolated atomic densities were computed for constructing the basis functions used in the LISA and GMBIS models.
Table~\ref{tab:charges} lists the isolated atomic densities calculated at the PBE/6-311+G(d,p) level of theory.
For hydrogen, atomic densities with charges of 0, -1, and -2 were computed.
For lithium, atomic densities with charges of +2, +1, 0, -1, and -2 were computed.
For boron, carbon, nitrogen, oxygen, fluorine, silicon, sulfur, chlorine, and bromine, atomic densities with charges of +3, +2, +1, 0, -1, and -2 were computed.
All DFT calculations were performed using the GAUSSIAN16 package.\cite{g16}
Input files for the atomic density calculations were generated using the \texttt{Horton 3} package.\cite{Chan2024}

\begin{table}
    \centering
    \caption{Atomic data with various charges employed in this work for fitting hyperparameters of the GMBIS model.}
    \label{tab:charges}
    \begin{tabular}{ll}
        \toprule
        Atom & Charges \\
        \midrule
        H & 0, -1, -2 \\
        Li & +2, +1, 0, -1, -2 \\
        B, C, N, O, F, Si, S, Cl, Br & +3, +2, +1, 0, -1, -2 \\
        \bottomrule
    \end{tabular}
\end{table}

To systematically assess the effects of basis function composition and parameter optimization strategies on partitioning performance, we designed a series of ISA-based models guided by two working hypotheses.
First, increasing the number of Slater basis functions should enhance basis-set completeness and thereby reduce the KL divergence.
Second, combining Gaussian basis functions with a Slater basis may improve the asymptotic accuracy of the electron density tails without compromising accuracy in the near-nuclear region.
These hypotheses motivated the structured development of the NLIS and LISA variants, incorporating controlled variations in the number and type of basis functions, as well as in the choice between fixed and optimized exponents.

Consequently, ISA and several of its variants were considered in this work, including six LISA models, four NLIS models, and two special NLIS variants: MBIS and GMBIS.
The first LISA model, referred to as LISA1, is the LISA model with Gaussian basis functions, as defined in Ref.~\citenum{Cheng2024} employing the \m{SC} solver.
The second LISA model, denoted by LISA2, corresponds to the LISA model with Gaussian basis functions and an additional Slater function for hydrogen, as defined in Ref.~\citenum{Cheng2024}, also using the \m{SC} solver.
Only results for molecules containing hydrogen thus differ between LISA1 and LISA2.

The MBIS model~\cite{Verstraelen2016} was used with convergence criteria set to match those of LISA1 and LISA2, as described in Ref.~\citenum{Cheng2024}.
The order $n_{ak}$ for the GMBIS model for each element was optimized using the NLIS model, with the same $k$ for each element as in MBIS.
The spherical averages of isolated atomic densities with various charges, as listed in Table~\ref{tab:charges}, were fitted using NLIS by selecting a set of hyperparameters $n_{ak}$.
The final hyperparameters were determined by minimizing the sum of KL divergences between the fitted NLIS pro-atom densities and the reference spherical densities.
For instance, the optimal $n_{ak}$ values for hydrogen were obtained by solving
\begin{align}
    \argmin_{n_{ak} \in [1, 2]} \left\{ \sum_{q=-2}^{0} \int_{\mathbb{R}^+} \langle \rho_q \rangle_s(r) \ln \left( \frac{\langle \rho_q \rangle_s(r)}{\rho_q^0(r; n_{ak})} \right) \, dr \right\},
\end{align}
where $\rho_q^0(r; n_{ak})$ denotes the pro-atom density constructed using the NLIS model with hyperparameters $n_{ak}$, and $\langle \rho_q \rangle_s(r)$ is the corresponding reference spherical density of the hydrogen atom with charge $q$.
The final values of $n_{ak}$ for each shell $k$ of each element are summarized in Table~\ref{tab:n_ak}.

\begin{table}[h!]
\centering
\caption{Optimized $n_{ak}$ values for each element used in the GMBIS model.}
\label{tab:n_ak}
\begin{tabular}{cccc}
\toprule
\textbf{Atom} & $k=1$ & $k=2$ & $k=3$ \\
\midrule
H  & 1.038 & $--$  & $--$  \\
Li & 1.000 & 1.069 & $--$  \\
B  & 1.069 & 1.000 & $--$  \\
C  & 1.138 & 1.000 & $--$  \\
N  & 1.172 & 1.000 & $--$  \\
O  & 1.172 & 1.000 & $--$  \\
F  & 1.138 & 1.000 & $--$  \\
Si & 1.111 & 1.667 & 1.000 \\
S  & 1.000 & 2.000 & 1.000 \\
Cl & 1.111 & 2.000 & 1.000 \\
Br & 1.111 & 2.000 & 1.000 \\
\bottomrule
\end{tabular}
\end{table}

The first general NLIS model, denoted NLIS1, uses the same number of basis functions as LISA1, with Gaussian functions ($n_{a,k}=2$) for all basis functions of an atom.
The optimized results of the NLIS model depend on the initial values~\cite{Verstraelen2016,Benda2022}.
In this work, the initial values for $c_{a,k}$ in all generalized NLIS models were set to $Z/N_k$, where $Z$ is the atomic number of atom $a$ and $N_k$ is the number of basis functions.
The initial values for $\alpha_{a,k}$ were set to those used in the MBIS model, except for the outermost shell of an atom, where a value of 0.5 was used instead of 2, as in the MBIS model~\cite{Verstraelen2016}.
The \m{SC} solver, described in Ref.~\citenum{Verstraelen2016}, was used for all NLIS models.
The second NLIS model, denoted NLIS2, was defined similarly to NLIS1, except that one Gaussian function was replaced by a Slater function for each element.
The third NLIS model (NLIS3) extends NLIS2 by replacing an additional Gaussian basis function with a Slater function for each atom.
The final generalized NLIS model (NLIS4) uses only Slater basis functions, with $n_{a,k}=1$ for all atoms.
The same initial values as in NLIS1 were used.

The third LISA model (LISA3) uses Gaussian basis functions fitted using the NLIS1 model.
Specifically, the NLIS1 model was applied to isolated atoms with various charges.
Data for atoms with charges of +2, +1, 0, -1, and -2 were considered, while for hydrogen, only charges of 0, -1, and -2 were used.
The optimized $\alpha_{a,k}$ and $c_{a,k}$ values were collected to form a new basis set for fitting molecular densities.
Functions with $c_{a,k}$ values smaller than $10^{-4}$ were removed.
Two functions were combined into one if $\frac{\alpha_1 - \alpha_2}{\alpha_1 + \alpha_2} < 0.1$, with the new $\alpha$ calculated as the average of $\alpha_1$ and $\alpha_2$.
The new $c_{a,k}$ was set as the sum of the $c_{a,k}$ values from the two functions.
The \m{M-NEWTON} solver was used.\cite{Cheng2024}
The fourth LISA model (LISA4) is based on LISA3, except that the \m{SC} solver was used.\cite{Cheng2024}
The fifth LISA model (LISA5) is similar to LISA3 but uses the NLIS4 model to fit atomic densities.
The final LISA model (LISA6) is the same as LISA5, except that the \m{SC} solver was used.\cite{Cheng2024}

Table~\ref{tab:methods} summarizes all methods used in this work.
An alternating iteration scheme, as described in Ref.~\citenum{Cheng2024}, was applied to all methods.
The criterion for the outer iteration, set to $10^{-8}$, follows the approach in Ref.~\citenum{Cheng2024}, while the criterion for the inner iteration was set to $10^{-9}$.

{
\footnotesize
\begin{longtable}{lccccc}
\caption{Summary of the partitioning models considered in this work.
These models differ in the type and origin of basis functions, the use of Slater-type functions, and whether parameters are fitted to atomic or molecular electron densities.
Here, $\rho_a$ denotes isolated atomic electron densities, and $\rho_\text{mol}$ denotes the total molecular electron density.
LS and KL refer to least-squares and Kullback-Leibler divergence fitting objectives, respectively.
All NLIS models employ a common optimization strategy but differ in the fixed number and type of Gaussian and Slater basis functions assigned to each atom.}
\label{tab:methods} \\
\toprule
\textbf{Model} & \textbf{Basis type} & \textbf{Parameter origin} & \textbf{Slater use} & \textbf{Solver} & \textbf{Ref.} \\
\midrule
\endfirsthead
\caption[]{(continued)} \\
\toprule
\textbf{Model} & \textbf{Basis type} & \textbf{Parameter origin} & \textbf{Slater use} & \textbf{Solver} & \textbf{Ref.} \\
\midrule
\endhead
\midrule
\multicolumn{6}{r}{\textit{Continued on next page}} \\
\endfoot
\bottomrule
\endlastfoot

\multicolumn{6}{l}{\textbf{(1) ISA, MBIS, and GMBIS models}} \\
ISA     & Numerical values         & $\rho_a$ on radial grid points          & No     & SC        & \citenum{C.Lillestolen2008,Lillestolen2009} \\
MBIS    & Slater                   & Fitted to $\rho_\text{mol}$            & No     & SC        & \citenum{Verstraelen2016,Cheng2024} \\
GMBIS   & Mixed, $1.0 \leq p \leq 2.0$ & Fitted to $\rho_a$ (KL)           & Yes    & SC        & This work \\[0.5ex]

\multicolumn{6}{l}{\textbf{(2) LISA models with LS-fitted atomic basis}} \\
LISA1   & Gaussian                 & Fitted to $\rho_a$ (LS)                & No     & SC        & \citenum{Cheng2024} \\
LISA2   & Gaussian + Slater       & Fitted to $\rho_a$ (LS)                & H only & SC        & \citenum{Cheng2024} \\[0.5ex]

\multicolumn{6}{l}{\textbf{(3) NLIS models with fixed basis types}} \\
NLIS1   & Gaussian                & Fitted to $\rho_\text{mol}$            & No         & SC        & This work \\
NLIS2   & Gaussian + Slater       & Fitted to $\rho_\text{mol}$            & 1 per atom & SC        & This work \\
NLIS3   & Gaussian + Slater       & Fitted to $\rho_\text{mol}$            & 2 per atom & SC        & This work \\
NLIS4   & Slater                  & Fitted to $\rho_\text{mol}$            & All        & SC        & This work \\[0.5ex]

\multicolumn{6}{l}{\textbf{(4) LISA models with NLIS-derived atomic basis}} \\
LISA3   & Gaussian                & NLIS1-fitted $\rho_a$                  & No     & M-NEWTON  & This work \\
LISA4   & Gaussian                & NLIS1-fitted $\rho_a$                  & No     & SC        & This work \\
LISA5   & Slater                  & NLIS4-fitted $\rho_a$                  & All    & M-NEWTON  & This work \\
LISA6   & Slater                  & NLIS4-fitted $\rho_a$                  & All    & SC        & This work \\
\end{longtable}
}

    \section{Results}
    \label{sec:res}
    \subsection{Partitioning accuracy and density reconstruction performance}
\label{ssec:numerical}
The optimized $n_{a,k}$, $\alpha_{a,k}$, and $c_{a,k}$ parameters for all test molecules across all methods are provided in Tables~S16--S30 in Section S2 of the Supporting Information.
Partitioning results, including $S$, $T$, and other quantities of interest for each atom in each molecule, are presented in Tables~S31--S45 in Section S3 of the Supporting Information.
Comparisons of radial densities $\rho_a^0(r)$ with $\langle \rho_a \rangle_s(r)$, as well as $\rho_a(r)$ with divergence $s_a(r)$ for atom $a$ in all test molecules across the NLIS1, LISA4, MBIS, and ISA methods, are shown in Figs. S1--S45 in Section S4 of the Supporting Information.

\begin{figure}[h]
    \centering
    \includegraphics[scale=0.7]{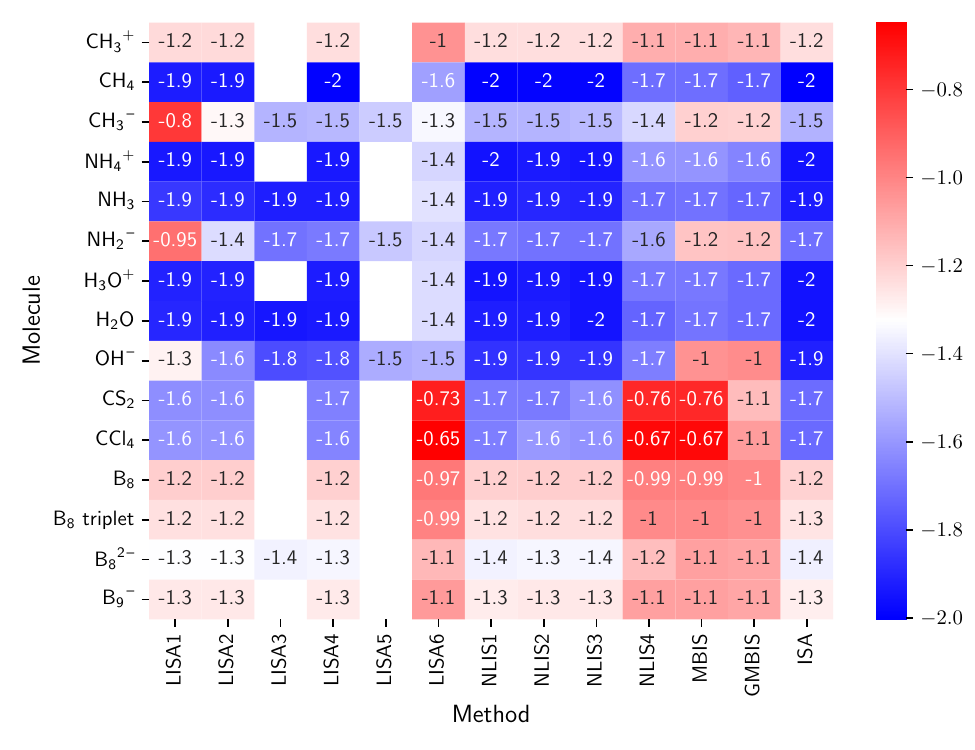}
    \caption{
         Heatmap of $\log_{10}(S/n_\text{atom})$ values for each molecule using various methods, where $n_\text{atom}$ is the number of atoms in the molecule.
        The color scale is applied globally across all entries: red indicates higher KL divergence (greater information loss), and blue indicates lower divergence.
        The colors reflect the relative representability of each molecule across all methods, not deviation from a specific reference model.
    }
    \label{fig:prop_s}
\end{figure}

Figure~\ref{fig:prop_s} presents the $\log_{10}(S/n_\text{atom})$ values for each molecule across various methods, where $n_\text{atom}$ denotes the number of atoms in the molecule.
The color scale is applied globally across all entries, with red representing higher KL divergence and blue representing lower divergence.
This visualization highlights which combinations of molecule and method result in greater information loss, rather than indicating deviation from a particular reference model.
Notably, some molecules such as \ce{CH3-} and \ce{NH2-} consistently exhibit higher $S$ values, suggesting that they are intrinsically more challenging to model.

\begin{figure}[h]
    \centering
    \includegraphics[scale=1.0]{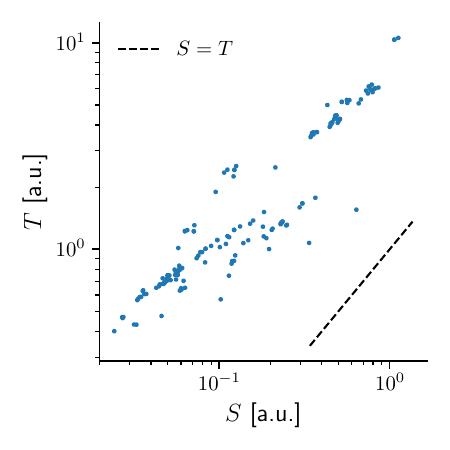}
    \caption{
Comparison between the molecular absolute KL divergence ($T$), defined in Eq.~\ref{eq:mol_abs_kl}, and the KL divergence ($S$), defined in Eq.~\ref{eq:divergence}, for all test molecules across different methods (both in a.u.).
    }
    \label{fig:prop_comp}
\end{figure}

The corresponding $\log_{10}(T/n_\text{atom})$ values are shown in Fig. S46 of the Supporting Information.
Results for LISA3 and LISA5 are unavailable for most molecules due to the large number of basis functions used, which frequently lead to invalid line searches in the \texttt{M-NEWTON} solver.\cite{Cheng2024}
A comparison of $T$ and $S$ across all test molecules and methods is provided in Fig.~\ref{fig:prop_comp}.
In general, methods with lower $S$ tend to exhibit lower $T$, although exceptions exist, as documented in Table S46 of the Supporting Information.
This indicates that the low $S$ values observed for the proposed methods are not simply the result of cancellation between large positive and negative contributions.
Therefore, for simplicity, we use $S$ as the primary criterion in the subsequent discussion.

\begin{figure}[h]
    \centering
    \includegraphics[scale=0.7]{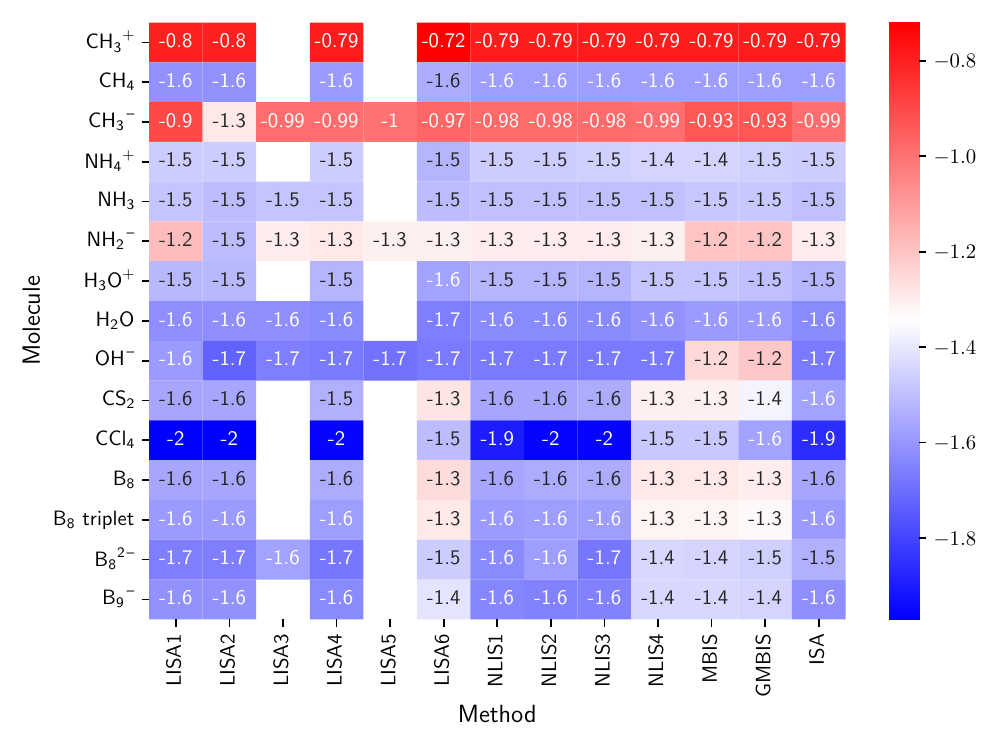}
    \caption{
        $\log_{10}(s^\text{I}_a)$ values for atom $a$ in each molecule, calculated using different methods.
        Here, $s_a^\text{I}$ measures the KL divergence between the full atomic density $\rho_a(\br)$ and its spherical average $\langle \rho_a \rangle_s(r)$, and thus reflects the degree of anisotropy.
        For molecules without a boron atom, $a$ corresponds to a chemically central atom (e.g., C, N, or O); for boron wheel molecules, $a$ refers to the central B atom.
    }
    \label{fig:prop_s1}
\end{figure}

\begin{figure}[h]
    \centering
    \includegraphics[scale=0.7]{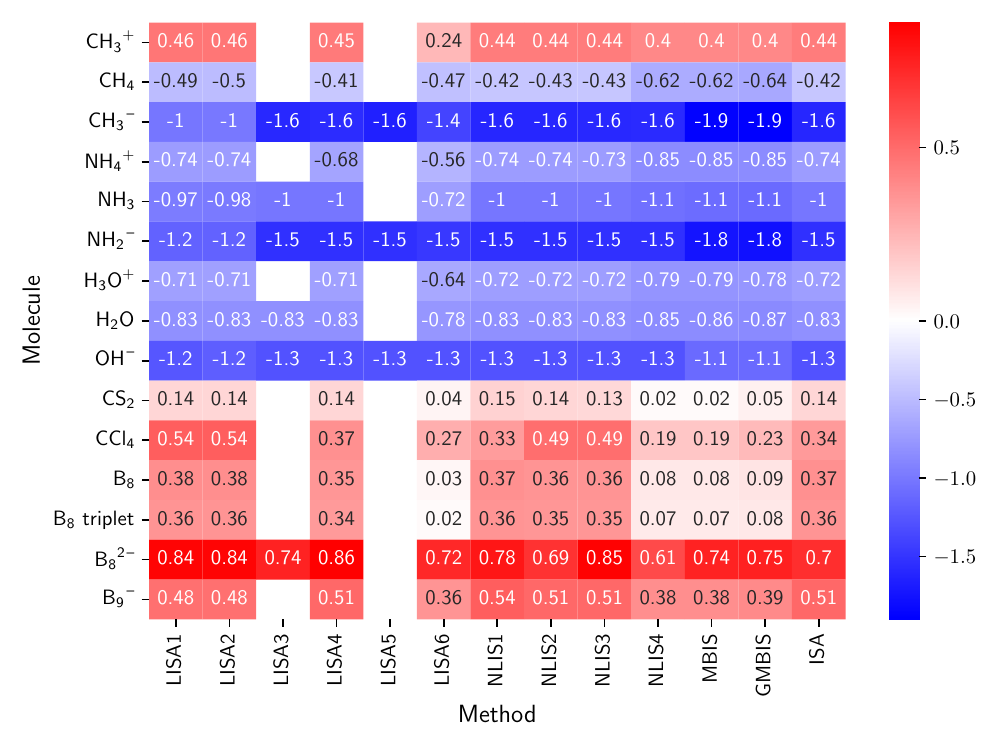}
    \caption{
        Atomic charges $q_a$ for atom $a$ in each molecule, calculated using different methods.
        For molecules without a boron atom, $a$ corresponds to a chemically central atom (e.g., C, N, or O), while for boron wheel molecules, $a$ refers to the central B atom.
        The color scale is applied globally across all molecule–method pairs, with red indicating more positive charges and blue indicating more negative values.
    }
    \label{fig:prop_q}
\end{figure}

Additionally, Fig.~\ref{fig:prop_s1} presents the $\log_{10}(s_a^\text{I})$ values for a selected atom $a$ in each molecule across different methods.
As defined in Eq.~\eqref{eq:atomic_divergence_part_1}, $s_a^\text{I}$ quantifies the KL divergence between the full three-dimensional AIM density $\rho_a(\br)$ and its spherical average $\langle \rho_a \rangle_s(r)$, thereby serving as a measure of the anisotropy of the atomic density.
For molecules without boron atoms, $a$ corresponds to a chemically central atom (typically C, N, or O), while for boron wheel molecules, $a$ refers to the central boron atom.
The figure allows comparison of how different ISA variants treat anisotropy in AIM densities across various systems.
Figures S47 and~\ref{fig:prop_q} are analogous to Fig.~\ref{fig:prop_s1}, presenting the corresponding $\log_{10}(s_a^\text{II})$ values (reflecting basis set completeness) and atomic charges $q_a$, respectively.

The ordering of $S$ for the NLIS models is NLIS1 $<$ NLIS2 $<$ NLIS3 $<$ NLIS4 $\leq$ MBIS, where $<$ ($\leq$) indicates that the method on the left produces a lower (or not higher) value than the method on the right.
Lower $S$ values generally correspond to better numerical performance in fitting atomic densities.
NLIS1 consistently achieves the lowest $S$ (Fig.~\ref{fig:prop_s}).
GMBIS produces $S$ values not greater than MBIS, though the differences are minor.
However, GMBIS does not fully address issues associated with MBIS, such as overly negative atomic charges in anionic molecules like \ce{CH3-} and \ce{NH3-}, as shown in Fig.~\ref{fig:prop_q}.

For most molecules, NLIS4 and MBIS yield similar $S$ values, and the number of unique exponents $\alpha_{a,k}$ obtained by NLIS4 matches the number of basis functions used in MBIS, as shown in Tables S11-S16 in the Supporting Information.
This suggests that simply adding more basis functions does not trivially improve MBIS results.
Additionally, LISA6 yields higher $S$ values compared to MBIS in certain molecular systems, as shown in Fig.~\ref{fig:prop_s}, despite using more Slater basis functions than MBIS.

For the LISA methods, the ordering is (LISA3 $\leq$) LISA4 $<$ LISA2 $\leq$ LISA1 $<$ (LISA5 $\leq$) LISA6, with parentheses indicating cases where the method has not converged.
The $S$ values for LISA4 closely align with NLIS1, demonstrating that Gaussian basis functions used in LISA4 are robust and generalizable.
Both LISA1 and LISA2 yield similar $S$ values to NLIS1, LISA4, or ISA, except for \ce{CH3-}, \ce{NH2-}, and \ce{OH-}, where LISA2 achieves a lower $S$ than LISA1 by including an additional Slater function for hydrogen.\cite{Cheng2024}
However, differences in atomic charges for C, N, and O atoms persist between LISA2 and LISA4 (or NLIS1 or ISA) for these molecules.
Adding an extra Slater function to hydrogen significantly reduces $s^\text{I}_\text{C}$ in \ce{CH3-} compared to MBIS, as shown in Fig.~\ref{fig:prop_s1}.
Nevertheless, this approach may still produce less reliable results than LISA4 or NLIS1 in chemical applications.

The $s_a^\text{I}$ values for selected atoms, shown in Fig.~\ref{fig:prop_s1}, indicate that LISA4, NLIS1, and NLIS2 align closely with ISA.
These methods yield similar atomic charges, as depicted in Fig.~\ref{fig:prop_q}.
Notably, when MBIS and GMBIS predict $s_a^\text{I}$ values close to those from LISA4, NLIS1, and ISA, the corresponding atomic charges are also similar.
However, for certain molecules, such as \ce{CH3-}, \ce{OH-}, \ce{NH2-}, \ce{CS2}, \ce{CCl4}, and the boron wheel molecules, MBIS and GMBIS produce significantly different $s_a^\text{I}$ values.
In these cases, the atomic charges predicted by MBIS and GMBIS deviate substantially from those obtained by LISA4 or NLIS1.

\begin{figure}[h]
    \centering
    \includegraphics[scale=0.7]{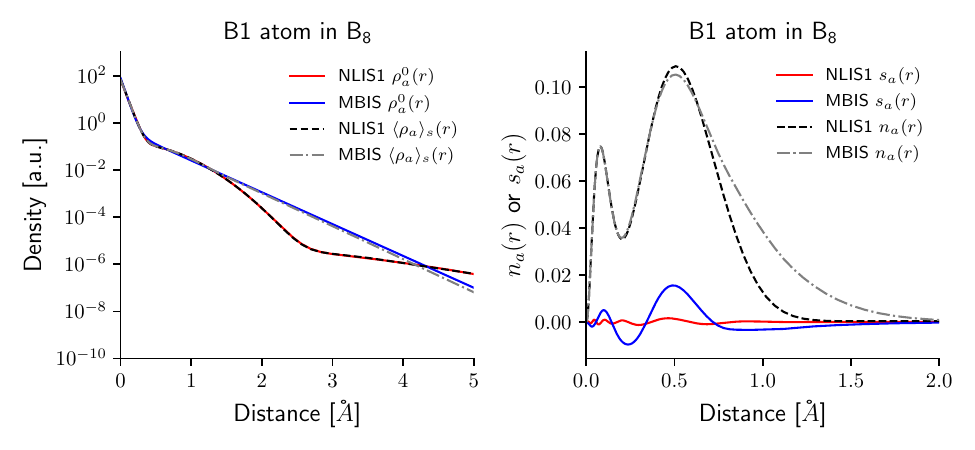}
    \includegraphics[scale=0.7]{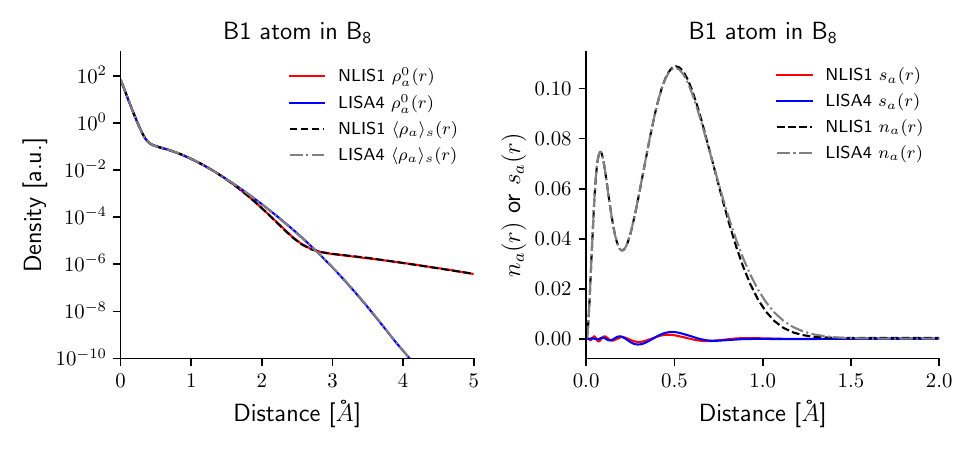}
    \includegraphics[scale=0.7]{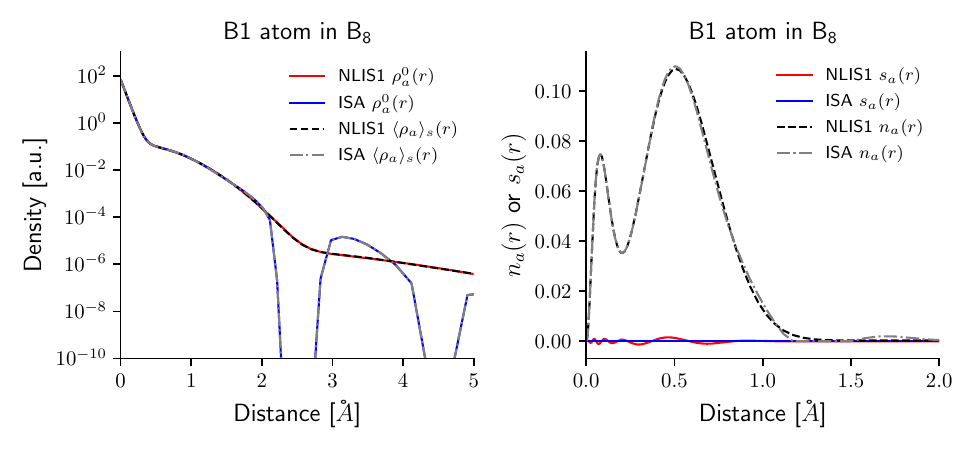}
    \caption{
    Comparison of quantities between NLIS1 and other methods, including MBIS, LISA4, and ISA, for the central B atom in \ce{B8}.
    The left panels show $\rhoa^0(r)$ and $\langle \rhoa \rangle_s(r)$, while the right panels display $s_a(r)$ and $n_a(r)$.
    }
    \label{fig:b8_nlis1_mbis}
\end{figure}

Delocalized systems exhibit substantial differences in atomic charges between MBIS or GMBIS and NLIS models.
For instance, the central B atom charge in \ce{B8} is 0.37 for NLIS1, compared to 0.08 and 0.09 for MBIS and GMBIS, respectively.
Fig.~\ref{fig:b8_nlis1_mbis} compares NLIS1 with MBIS, LISA4, and ISA methods for the central B atom in \ce{B8}.
The left panels show $\rhoa^0(r)$ and $\langle \rhoa \rangle_s(r)$, while the right panels display $s_a(r)$ and $n_a(r)$.
Compared to MBIS, NLIS1 shows a less smooth decay in both $\rho_a^0(r)$ and the spherical average $\langle \rho_a \rangle_s(r)$, particularly in the tail region.
However, $s_a(r)$ from NLIS1 exhibits much smaller oscillation amplitudes, resulting in a lower $s^\text{II}_a$ value.
LISA4 partially agrees with NLIS1, especially in the core region, but deviations appear in the density tail, due to the use of fewer diffuse basis functions.
Compared to ISA, NLIS1 reproduces the overall trends in both $\rho_a^0(r)$ and $\langle \rho_a \rangle_s(r)$ without introducing discontinuities.
These observations indicate that NLIS1 yields mathematically well-behaved partitioning functions, with the KL divergence applied as the objective functional.

\subsection{Assessment of computational robustness and electrostatic accuracy}
\label{ssec:chemical}

In this section, we evaluate the chemical performance of different partitioning methods in terms of computational robustness and their ability to reproduce the electrostatic potential (ESP).
For simplicity, we consider four representative models: LISA4, MBIS, NLIS1, and ISA.
We adopt the same molecular test set as in Ref.~\citenum{Heidar-Zadeh2024}, including \ce{CH3+}, \ce{CH4}, \ce{CH3-}, \ce{NH4+}, \ce{NH3}, \ce{NH2-}, \ce{H3O+}, \ce{H2O}, and \ce{OH-}.

To assess computational robustness, we examine the dependence of AIM charges on the level of theory at a fixed molecular geometry.
Calculations were performed using four XC functionals (HF, LDA, PBE, and B3LYP) in combination with 12 basis sets from the (d-aug)-cc-pVXZ family (X = D, T, Q, 5); see Section~\ref{sec:details} for further information.
This setup enables a comprehensive evaluation of the sensitivity of each model to both the choice of XC functional and basis set.
Atomic charges on C, N, and O atoms were computed for all combinations to assess numerical stability and chemical consistency across methods.

To evaluate each model's ability to reproduce the ESP, we adopt the approach proposed by Farnaz \textit{et al.} in Ref.~\citenum{Heidar-Zadeh2024}.
The root-mean-square error (RMSE) quantifies the deviation between the reference ESP and that reconstructed from the AIM charges.
As a reference, we also include the Hu–Lu–Yang charge fitting method, based on Gaussian’s standard atomic densities (HLYGAt),\cite{Hu2007a} which directly fits atomic charges to the ESP and thus serves as a baseline for benchmarking AIM-derived models.

\begin{figure}[h]
    \centering
    \includegraphics[scale=0.7]{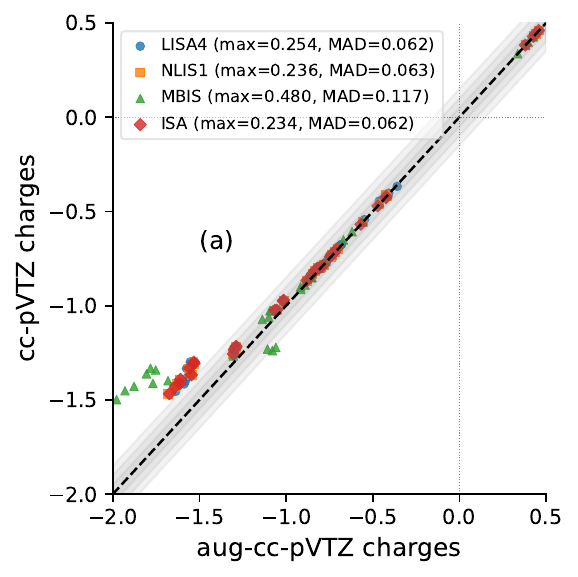}
    \includegraphics[scale=0.7]{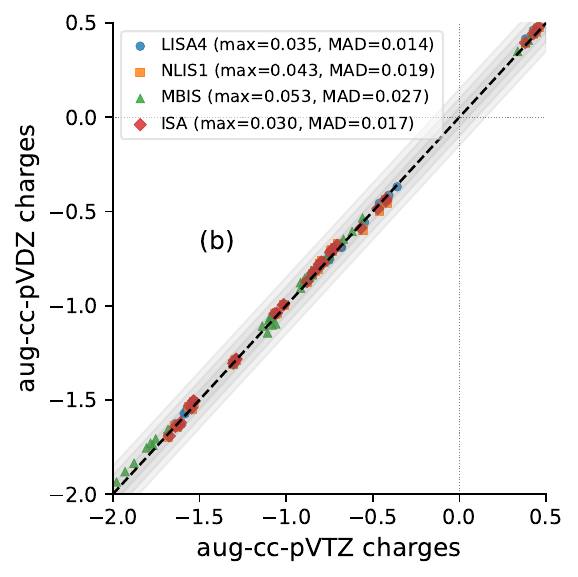}\\
    \includegraphics[scale=0.7]{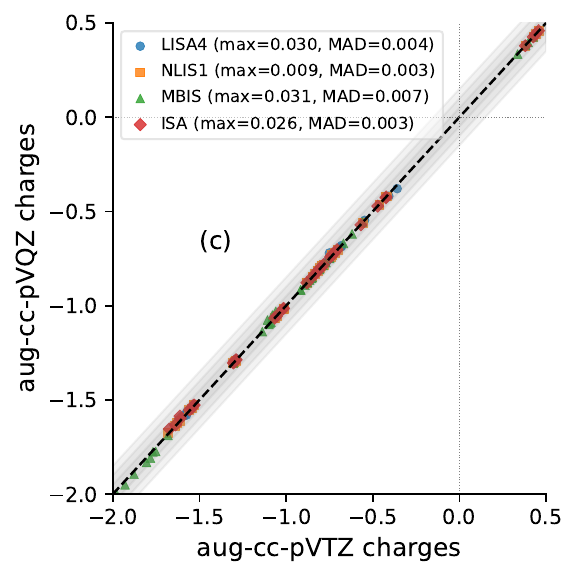}
    \includegraphics[scale=0.7]{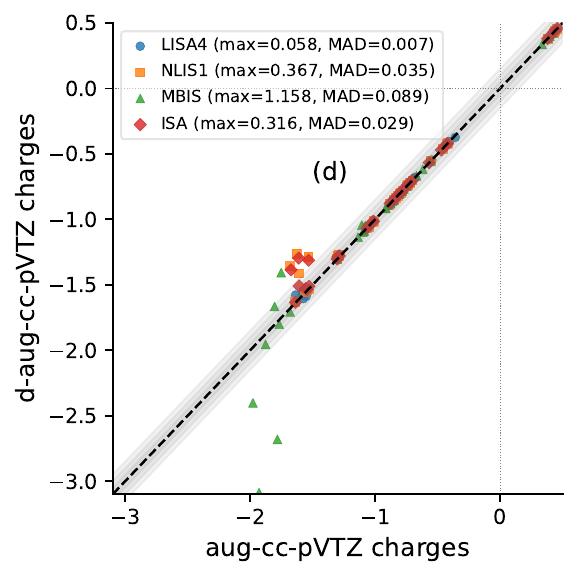}
    \caption{
        Computational robustness of atomic charges $q_\text{C}$, $q_\text{N}$, $q_\text{O}$ from the \ce{CH3+}, \ce{CH4}, \ce{CH3-}, \ce{NH4+}, \ce{NH3}, \ce{NH2-}, \ce{H3O+}, \ce{H2O}, and \ce{OH-} when changing the basis set from aug-cc-pVTZ to cc-pVTZ (a), aug-cc-pVDZ (b), aug-cc-pVQZ (c), and d-aug-cc-pVTZ (d) while keeping the theory fixed.
        ``max'' and ``MAD'' denote the maximum deviation and the mean absolute deviation, respectively, between the two basis sets.
        The equally distant lines around y = x denote the 0.05, 0.10, 0.15 margins.
        Similar plots for other basis sets (Fig. S48) can be found in the Supporting Information.
    }
    \label{fig:basis_robustness}
\end{figure}

\begin{figure}[h]
    \centering
    \includegraphics[scale=0.7]{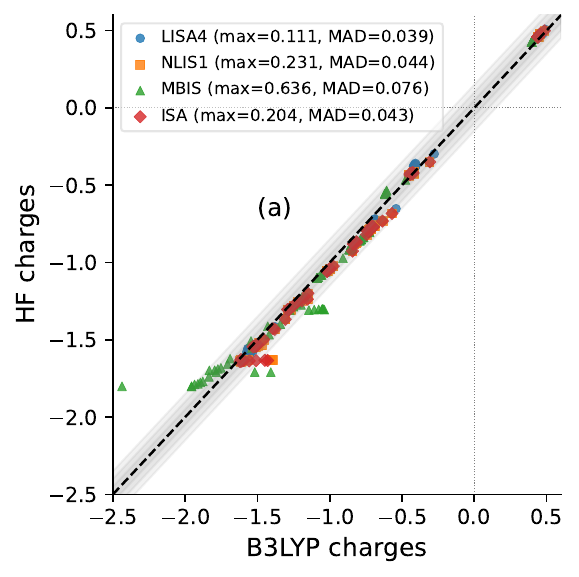}
    \includegraphics[scale=0.7]{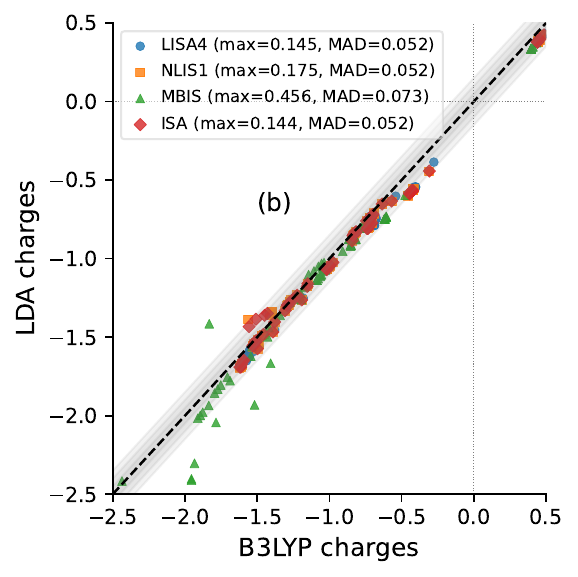}
    \includegraphics[scale=0.7]{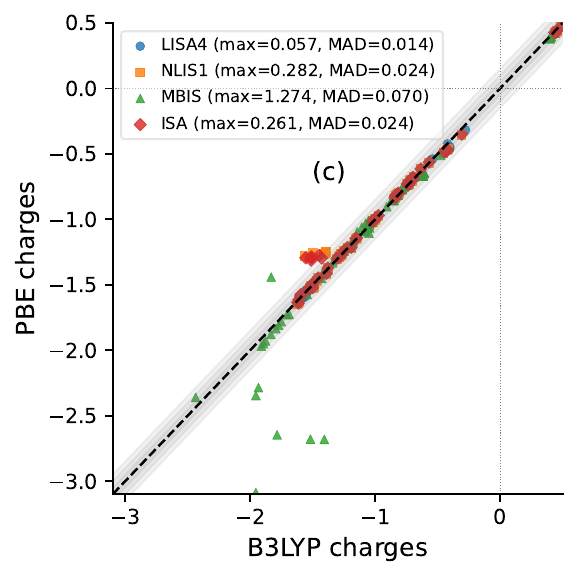}
    \caption{
        Computational robustness of atomic charges $q_\text{C}$, $q_\text{N}$, $q_\text{O}$ from the \ce{CH3+}, \ce{CH4}, \ce{CH3-}, \ce{NH4+}, \ce{NH3}, \ce{NH2-}, \ce{H3O+}, \ce{H2O}, and \ce{OH-} when changing the level of theory from B3LYP to HF (a), LDA (b), PBE (c) while keeping the basis set fixed.
        ``max'' and ``MAD'' denote the maximum deviation and the mean absolute deviation, respectively, between the two XC functionals.
        The equally distant lines around y = x denote the 0.05, 0.10, 0.15 margins.
    }
    \label{fig:xc_robustness}
\end{figure}

Figures~\ref{fig:basis_robustness} and~\ref{fig:xc_robustness} illustrate the computational robustness of atomic charges $q_\text{C}$, $q_\text{N}$, and $q_\text{O}$ with respect to variations in the basis set and the XC functional, respectively.
In Fig.~\ref{fig:basis_robustness}, aug-cc-pVTZ serves as the reference, and the computed charges are compared to those obtained using cc-pVTZ (a), aug-cc-pVDZ (b), aug-cc-pVQZ (c), and d-aug-cc-pVTZ (d).
Additional comparisons involving other basis sets are provided in the Supporting Information (Fig.~S48).
The original numerical values are also tabulated in Table~S47 of the Supporting Information.

Among all models, LISA4 exhibits the highest robustness, as evidenced by consistently lower maximum absolute deviations and mean absolute deviations (MADs).
This trend persists across the other basis sets, confirming the stability of LISA4 with respect to basis set variation.
Notably, the largest deviations are observed when comparing aug-cc-pVTZ with cc-pVXZ basis sets (X = D, T, Q, 5), as shown in Fig.~\ref{fig:basis_robustness}(a) and Fig.~S48 of the Supporting Information, particularly for anionic species.
This behavior may be attributed to the absence of diffuse augmentation functions in the cc-pVXZ sets, which limits their ability to accurately describe the slowly decaying electron densities typical of anions.

As shown in Fig.~\ref{fig:basis_robustness}(d), the use of more diffuse basis sets results in larger MADs for MBIS, NLIS1, and ISA, primarily due to challenges in modeling anionic systems.
In contrast, LISA4 remains robust across all tested basis sets.
This robustness is likely due to its use of fixed basis functions, constructed from the density fitting of isolated atoms and ions at the PBE/6-311G** level.
Additional comparisons involving d-aug-cc-pVXZ basis sets are provided in the Supporting Information (Fig.~S48).

Figure~\ref{fig:xc_robustness} examines the sensitivity of AIM charges to the choice of XC functional, using B3LYP as the reference.
LISA4 again shows the highest robustness, followed closely by NLIS1 and ISA.
In contrast, MBIS exhibits significant fluctuations, especially in the comparison between B3LYP and PBE as shown in Fig.~\ref{fig:xc_robustness}(c), where the maximum deviation exceeds 1.2 a.u.
This may be attributed to the fact that electron densities computed using local or semi-local functionals do not exhibit the correct asymptotic exponential decay, thereby reducing MBIS's consistency.

Tables~\ref{tbl:charge_c}--\ref{tbl:charge_o} summarize atomic charges for C, N, and O atoms computed at the B3LYP/aug-cc-pVTZ level for LISA4, MBIS, NLIS1, ISA, and HLYGAt.
The expected chemical trends are well reproduced by most models,\cite{Heidar-Zadeh2018} namely:
$q_\text{C}$(\ce{CH3+}) > $q_\text{C}$(\ce{CH4}) > $q_\text{C}$(\ce{CH3-}),
$q_\text{N}$(\ce{NH4+}) > $q_\text{N}$(\ce{NH3}) > $q_\text{N}$(\ce{NH2-}),
and $q_\text{O}$(\ce{H3O+}) > $q_\text{O}$(\ce{H2O}) > $q_\text{O}$(\ce{OH-}).
The charges and ESP values obtained using MBIS and HLYGAt are in good agreement with the values reported in Ref.~\citenum{Heidar-Zadeh2024}.
The small differences may arise from differences in molecular coordinates or the sampling points used in the ESP fitting.

All models reproduce the expected chemical trends.
Among them, MBIS is qualitatively consistent with these trends but tends to overestimate the charge magnitudes, particularly for anions such as \ce{CH3-} and \ce{NH2-}, where it yields the most negative charges across all models.
This observation is consistent with previous findings.\cite{Heidar-Zadeh2018,Heidar-Zadeh2024}

ESP fitting performance is quantified by the RMSE$_\text{ESP}$ values shown in each table.
As expected, HLYGAt consistently achieves the lowest RMSE due to its fitting-based design, except for \ce{CH3+}, where its RMSE$_\text{ESP}$ is slightly higher than those of the other models.
One possible explanation is the specific distribution or selection of grid points used in the ESP fitting procedure.

Among the AIM-based models, LISA4 strikes the best balance between ESP accuracy and chemical consistency, followed by NLIS1 and ISA.
By contrast, MBIS generally yields larger RMSEs than the other schemes, except for \ce{CH3+}, where it produces the lowest RMSE$_\text{ESP}$ among all models.

\begin{table}
    \centering
    \caption{Comparison of carbon atomic charges at B3LYP/aug-cc-pVTZ level of theory for the LISA1, MBIS, NLIS1, and modified HLYGAt methods.}
    \begin{tabular}{lrrrrrr}
\toprule
scheme & \multicolumn{2}{c}{\ce{CH3+}} & \multicolumn{2}{c}{\ce{CH4}} & \multicolumn{2}{c}{\ce{CH3-}} \\
 & $q_\text{C}$ & RMSE$_{\mathrm{ESP}}$ & $q_\text{C}$ & RMSE$_{\mathrm{ESP}}$ & $q_\text{C}$ & RMSE$_{\mathrm{ESP}}$ \\
\cmidrule(lr){2-3} \cmidrule(lr){4-5} \cmidrule(lr){6-7}
LISA4 & 0.45 & 1.5 & -0.41 & 0.7 & -1.57 & 7.5 \\
MBIS & 0.40 & 1.5 & -0.62 & 0.4 & -1.88 & 7.8 \\
NLIS1 & 0.44 & 1.5 & -0.42 & 0.7 & -1.61 & 7.5 \\
ISA & 0.44 & 1.5 & -0.42 & 0.7 & -1.61 & 7.5 \\
HLYGAt & 0.54 & 1.6 & -0.56 & 0.5 & -1.49 & 7.7 \\
\bottomrule
\end{tabular}
     \label{tbl:charge_c}
\end{table}

\begin{table}
    \centering
    \caption{
        Same as Table~\ref{tbl:charge_c} but for N atom.
}
    \begin{tabular}{lrrrrrr}
\toprule
scheme & \multicolumn{2}{c}{\ce{NH4+}} & \multicolumn{2}{c}{\ce{NH3}} & \multicolumn{2}{c}{\ce{NH2-}} \\
 & $q_\text{N}$ & RMSE$_{\mathrm{ESP}}$ & $q_\text{N}$ & RMSE$_{\mathrm{ESP}}$ & $q_\text{N}$ & RMSE$_{\mathrm{ESP}}$ \\
\cmidrule(lr){2-3} \cmidrule(lr){4-5} \cmidrule(lr){6-7}
LISA4 & -0.68 & 0.4 & -1.02 & 2.9 & -1.54 & 6.8 \\
MBIS & -0.85 & 0.2 & -1.09 & 3.4 & -1.75 & 9.2 \\
NLIS1 & -0.74 & 0.3 & -1.02 & 2.9 & -1.53 & 6.7 \\
ISA & -0.74 & 0.3 & -1.02 & 2.9 & -1.53 & 6.7 \\
HLYGAt & -0.81 & 0.2 & -0.87 & 2.5 & -1.30 & 6.4 \\
\bottomrule
\end{tabular}
     \label{tbl:charge_n}
\end{table}

\begin{table}
    \centering
    \caption{
    Same as Table~\ref{tbl:charge_c} but for O atom.
}
    \begin{tabular}{lrrrrrr}
\toprule
scheme & \multicolumn{2}{c}{\ce{H3O+}} & \multicolumn{2}{c}{\ce{H2O}} & \multicolumn{2}{c}{\ce{OH-}} \\
 & $q_\text{O}$ & RMSE$_{\mathrm{ESP}}$ & $q_\text{O}$ & RMSE$_{\mathrm{ESP}}$ & $q_\text{O}$ & RMSE$_{\mathrm{ESP}}$ \\
\cmidrule(lr){2-3} \cmidrule(lr){4-5} \cmidrule(lr){6-7}
LISA4 & -0.71 & 2.9 & -0.83 & 3.4 & -1.29 & 4.3 \\
MBIS & -0.79 & 3.4 & -0.86 & 3.9 & -1.11 & 5.4 \\
NLIS1 & -0.72 & 2.9 & -0.83 & 3.4 & -1.29 & 4.3 \\
ISA & -0.72 & 2.9 & -0.83 & 3.4 & -1.29 & 4.3 \\
HLYGAt & -0.45 & 1.9 & -0.68 & 2.3 & -1.18 & 4.1 \\
\bottomrule
\end{tabular}
     \label{tbl:charge_o}
\end{table}

\subsection{Connection to the AVH scheme}
\label{ssec:conn_to_avh}

In this section, we explore the connection between the present model and the AVH model.\cite{Heidar-Zadeh2024}
The AVH model has recently been proposed as a promising alternative for atomic partitioning in the chemical community.
Its basis functions are constructed from the promolecular density as a convex linear combination of electron densities corresponding to selected charge states of isolated atoms and ions, a formulation that is conceptually aligned with the LISA4 framework.

In LISA4, atomic densities are represented using Gaussian-type basis functions, with hyperparameters optimized via the NLIS1 model.
In contrast, the AVH model directly employs spherically averaged atomic densities without expanding them into Gaussian primitives.
Spline interpolation is then used to extend these radial density functions smoothly across molecular space.

The spherically averaged electron density of an isolated atom with charge $q$ can be approximated as
\begin{align}
    \langle \rho_{q; a} \rangle_s (r) \approx \rho_{q; a}^0(r) = \sum_{q; ak} c_{q; ak} \exp(-\alpha_{q; ak} r^2),
    \label{eq:avh_basis}
\end{align}
where $c_{q; ak}$ and $\alpha_{q; ak}$ are coefficients determined by the NLIS1 optimization, and $k$ indexes the Gaussian primitives.

In the LISA4 model, all primitive functions $\exp(-\alpha_{q; ak} r^2)$ associated with the selected charge states $q$ are pooled to form the basis set for element $a$.
At this stage, linear dependencies among these primitives are not explicitly removed; they are handled during the construction of the practical LISA4 basis set (see Section~\ref{sec:details} for implementation details).

Within this framework, the basis function in an AVH-like model corresponds exactly to the contracted form given in Eq.~\eqref{eq:avh_basis}, namely $\rho_{q; a}^0(r)$, which is a fixed linear combination of Gaussian primitives from the LISA4 model, with contraction coefficients $c_{q; ak}$ determined by the NLIS1 optimization.
Thus, AVH-like behavior can be reproduced by fixing the contraction coefficients and selecting appropriate charge states $q$ for each element.

Another important distinction between LISA4 and AVH lies in how the basis functions are generated and applied.
In LISA4, the Gaussian basis functions are optimized once at the PBE/6-311G** level and are used consistently for all calculations, independent of the electronic structure method used for the molecular density.
In contrast, the AVH model is inherently dependent on the level of theory used in the promolecular density construction, and thus its basis functions vary with changes in the underlying electronic structure calculation.
A more comprehensive comparison between LISA4 and AVH, along with an assessment of the LISA4 model in practical applications employing basis functions adapted to the level of theory, as done in AVH, will be undertaken in future work.
 
    \section{Summary}
    \label{sec:summary}
    This study investigates linear and non-linear approximations of ISA partitioning methods, denoted as LISA and NLIS, respectively, employing exponential basis functions as an alternative to the pointwise discretization used in the original ISA method.
A general NLIS model is formulated using a generalized Lagrangian framework with exponential basis functions, treating both the exponential order and degree as variables.
However, due to numerical difficulties, the degree of exponential functions is treated as a hyperparameter in all NLIS models.
The well-established MBIS model is reproduced by using only Slater-type basis functions with the number of basis functions fixed for each element.
A generalized MBIS model is developed by optimizing the degree of exponential basis functions for each element type independently, relaxing the constraint of using pure Slater-type forms.

LISA models are reproduced by fixing the exponents and degrees of the basis functions, and a set of LISA variants is proposed.
The LISA4 model employs Gaussian basis functions, with exponents optimized by fitting atomic densities using the NLIS model.
All computations were performed using the open-source \texttt{Horton-Part} program and benchmarked on molecules and charged ions, including four boron wheel molecules with significantly delocalized bonding environments.

To evaluate model performance, several partitioning metrics are introduced, including absolute divergence and the radial distribution of atomic divergence.
Among the methods tested, NLIS1, which exclusively uses Gaussian-type basis functions, achieved the lowest molecular divergence values and excelled in other metrics.
The results for the LISA4 model closely align with those of NLIS1 across all metrics.
Both LISA4 and NLIS1 show strong agreement with the ISA method, particularly in terms of atomic charges.

In summary, this work addresses the well-known issue in ISA partitioning by employing basis function approaches in all proposed models.
Considering computational efficiency, numerical stability, and the ability to reproduce electrostatic potentials, LISA4 emerges as a promising and balanced model.
Nevertheless, further validation against broader benchmarks, including geometry sensitivity and intermolecular electrostatics, will be pursued in future work.
 
    \begin{acknowledgement}
        Y.C. and B.S. acknowledge funding by the Deutsche Forschungsgemeinschaft (DFG, German Research Foundation) - Project number 442047500 through the Collaborative Research Center ``Sparsity and Singular Structures'' (SFB 1481).
We also thank the Deutsche Forschungsgemeinschaft (DFG, German Research Foundation) for supporting this work by funding - EXC2075 – 390740016 under Germany's Excellence Strategy.
We acknowledge the support by the Stuttgart Center for Simulation Science (SimTech).
The resources and services used in this work were provided by the VSC (Flemish Supercomputer Center), funded by the Research Foundation - Flanders (FWO) and the Flemish Government.
     \end{acknowledgement}

    \begin{suppinfo}
        The Supplementary Material is a PDF document that includes:
(1) the optimized coordinates of the 15 test molecules;
(2) the exponential basis functions used for each molecule;
(3) the partitioning results of different models for each molecule;
(4) the radial density and entropy of a selected atom in each molecule;
(5) $T$ values and cases where $S$ and $T$ values are inconsistent;
(6) atomic charges obtained at different levels of theory; and
(7) the robustness of atomic charges with respect to changes in the basis set.
     \end{suppinfo}

    \providecommand{\latin}[1]{#1}
\makeatletter
\providecommand{\doi}
  {\begingroup\let\do\@makeother\dospecials
  \catcode`\{=1 \catcode`\}=2 \doi@aux}
\providecommand{\doi@aux}[1]{\endgroup\texttt{#1}}
\makeatother
\providecommand*\mcitethebibliography{\thebibliography}
\csname @ifundefined\endcsname{endmcitethebibliography}
  {\let\endmcitethebibliography\endthebibliography}{}

\end{document}